\documentclass[letterpaper, aps, showpacs,onecolumn]{revtex4}

\usepackage{amsmath}
\usepackage{amssymb}
\usepackage{latexsym}
\usepackage{graphicx,epstopdf}
\usepackage{hyperref}

\begin{document}

\title{Calculation of eigenvalues by Greens-functions\\
and the Lippmann-Schwinger equation}
\author{Alexander Jurisch}
\affiliation{ajurisch@ymail.com, Munich, Germany}

\begin{abstract}
We calculate eigenvalues of one-dimensional quantum-systems by the exact numerical solution of the Lippmann-Schwinger equation, analogous to the scattering problem. To illustrate our method, we treat elementary problems: the harmonic and quartic oscillator, a symmetric and a skew double-well potential, and potentials with finite and infinite depth. Furthermore, we show how our method can be used for eigenvalue-engineering.

\end{abstract}
\maketitle
\section{Introduction}
The numerical calculation of the eigenvalues of quantum-systems is an elementary, but also a formidable task. There exist a multitude of methods, from which we shall only mention, without citations, the most well-known ones: the approach by Numerov, the Lanczos-algorithm, the discretized-matrix method (DM), or the shooting-method. All these approaches, and also all approaches we have not mentioned here, use highly elaborated numerical techniques that are tailor-made to treat differential-equations of the second order, but none of these approaches has a connection to the basic properties of the Schr\"odinger-equation.

Contrary to this, our present approach does not employ sophisticated numerical techniques. It makes direct use of the stationary Schr\"odinger-equation and it's integral-representation, which is the Lippmann-Schwinger equation. Furthermore, we use properties of the Greens-function and the symmetry of space, which is described by the solutions of the Laplace-equation. We remind the reader about the fact that the solutions of the $d$-dimensional Laplace-equation are the Gegenbauer-polynomials. A special case in three dimensions are the Legendre-polynomials, and thus the spherical harmonics. Consequently, the Gegenbauer-polynomials describe a multipole-expansion that is related to the symmetry of the $d$-dimensional space: monopole is symmetric, dipole is anti-symmetric, and so on. It is thus clear, that the one-dimensional Laplace-equation has solutions that are related to a monopole and a dipole.

In summary, our approach only uses very elementary ingredients.

To illustrate our method, we discuss the harmonic and the quartic oscillator. Furthermore, we treat a symmetric and a skew double-well potential, and the attractive $\cosh^{-2}$-potential as an example for systems with final depth. As examples for singular potentials we examine the Coulomb-problem and the van der Waals-potential tail. To regularize the singularity of the van der Waals-potential tail we employ the Duru-Kleinert transform \cite{Kleinert1, Kleinert2}. Finally, we demonstrate how our approach can be used for eigenvalue-engineering.

\section{The method}
The general solution of the Schr\"odinger-equation is given by the Lippmann-Schwinger equation
\begin{equation}
\psi({\bf{r}})\,=\,\phi({\bf{r}})\,+\,\int\,d{\bf{r}}'\,G\left(\left|{\bf{r}}\,-\,{\bf{r}}'\right|\right)\,U\left({\bf{r}}'\right)\,\psi({\bf{r}}')\quad,
\label{LippmannSchwinger}\end{equation}
with the potential $U$ and the Greens-function $G$. Eq. (\ref{LippmannSchwinger}) is usually applied to calculate scattering states, but as Eq. (\ref{LippmannSchwinger}) is the general solution of the Schr\"odinger-equation it also must contain bound-state solutions if the potential $U$ has bound states.

In the case of bound states, all functions that enter Eq. (\ref{LippmannSchwinger}) have to be real. In order to ensure this condition, we shall examine the Greens-function at first. The stationary Greens-function is a solution of the Helmholtz-equation
\begin{equation}
(\nabla^{2}\,+\,E)\,G\left(\left|{\bf{r}}\,-\,{\bf{r}}'\right|,\,E\right)\,=\,\delta({\bf{r}}\,-\,{\bf{r}}')\quad, 
\label{Helmholtz}\end{equation}
where $E$ is the energy. Thus, for $E\rightarrow0$ the Greens-function must be a solution of the Poisson-equation
\begin{equation}
\nabla^{2}\,G\left(\left|{\bf{r}}\,-\,{\bf{r}}'\right|,\,E\,=\,0\right)\,=\,\delta({\bf{r}}\,-\,{\bf{r}}')\quad,
\label{Poisson}\end{equation}
which implies that $G$ either has to be a real, or that the imaginary part must vanish. In one dimension, the Greens-functions are given by
\begin{eqnarray}
G\left(\left|{\bf{r}}\,-\,{\bf{r}}'\right|,\,E\right)&=&-\frac{i}{2\,\sqrt{E}}\,\exp\left[-\,i\,\sqrt{E}\,|x\,-\,x'|\right]\quad,\\
\lim_{E\rightarrow 0}\,G\left(\left|{\bf{r}}\,-\,{\bf{r}}'\right|,\,E\right)&=&\frac{1}{2}\,|x\,-\,x'|\quad.
\label{Greens1D}\end{eqnarray}
However, the limit $E\rightarrow 0$ is only fulfilled by the real part of $G\left(\left|x\,-\,x'\right|,\,E\right)$ since
\begin{equation}
\lim_{E\rightarrow 0}\,\mathrm{Re}\{G\left(\left|x\,-\,x'\right|,\,E\right)\}\,=\,\lim_{E\rightarrow 0}\,\frac{1}{2\,\sqrt{E}}\,\sin\left[\sqrt{E}\,|x\,-\,x'|\right]\,=\,\frac{1}{2}|x\,-\,x'|\quad.
\label{1DLimitPoisson}\end{equation}
The imaginary part of $G$ is given by a cosine and does not fulfill this requirement, as it diverges. Consequently, the imaginary part is of no interest here.

The next question we have to treat is the question of the \emph{incoming} state $\phi$. The only information about a bound state we know for sure is, that it has to fulfill a condition of symmetry. Functions that describe the symmetry of space are but given by the Laplace-equation $\nabla^{2}\,\phi=0$. In one dimension, the solution of the Laplace-equation reads
\begin{equation}
\phi(x)\,=\,1\,+\,x\,\quad.
\label{1DLaplace}\end{equation}
From this general solution we can extract a symmetric (monopole) solution $\phi_{\mathrm{s}}(x)=1$, and an anti-symmetric (dipole) solution $\phi_{\mathrm{as}}(x)=x$. Both these solutions certainly fulfill the possible symmetries that are present in a one-dimensional space. The solutions of the Laplace-equation act as projectors. All eigenvalues are present in the product of the Greens-function with the potential, $G\,U$, and the \emph{incoming} state $\phi$ filters them out. Thus, for the calculation of symmetric states $\phi_{\mathrm{s}}(x)=1$ has to be used as an \emph{incoming} state, while for anti-symmetric or odd states $\phi_{\mathrm{odd}}(x)=x$ has to be used.

The energy is generally calculated by
\begin{equation}
E\,=\,\left<\psi({\bf{r}}, E)|\hat{H}|\psi({\bf{r}}, E)\right>\quad.
\label{Energycondition1}\end{equation}
If now $E_{n}$ is an eigen-energy, we may also take into account variations up to the first order, giving
\begin{equation}
E_{n}\,=\,\left<\psi({\bf{r}}, E_{n})|\hat{H}|\psi({\bf{r}}, E_{n})\right>\,+\,\frac{\partial}{\partial\,E}\,\left<\psi({\bf{r}}, E)|\hat{H}|\psi({\bf{r}}, E)\right>{\Big{|}}_{E=E_{n}}\delta\,E\quad.
\label{eigenenergy11}\end{equation}
To ensure the uniqueness of the eigen-energy $E_{n}$ it should be stable under variations $\delta E$. Consequently, we must require that for an eigen-energy it must hold
\begin{equation}
\frac{\partial}{\partial\,E}\,\left<\psi({\bf{r}}, E)|\hat{H}|\psi({\bf{r}}, E)\right>{\Big{|}}_{E=E_{n}}\,=\,0\quad.
\label{Energycondition2}\end{equation}
Any eigenvalue has to fulfill the conditions that are given by Eqs. (\ref{Energycondition1}, \ref{Energycondition2}) simultaneously. Below we will see, that indeed both of these conditions must be satisfied in order to guarantee the existence and uniqueness of the eigenvalue.

Finally, the exact solution of the Lippmann-Schwinger equation Eq. (\ref{LippmannSchwinger}) by matrix-inversion
\begin{equation}
\psi_{\mu}\,=\,\left(\delta_{\mu\nu}\,-\,G_{\mu\nu}\,U_{\nu}\right)^{-1}\,\phi_{\nu}\quad
\label{solution1D}\end{equation}
is easily accessible.

\section{Examples}
We shall treat four examples. First, the harmonic and quartic oscillator. Second, a symmetric and a skew double-well potential, third the inverse and attractive $\cosh^{-2}$-potential and, finally, the Coulomb-problem and the van der Waals-potential tail.

\subsection{Harmonic and quartic oscillator}
\begin{figure}[t]\centering\vspace{-1.05cm}
\rotatebox{0.0}{\scalebox{1.}{\includegraphics{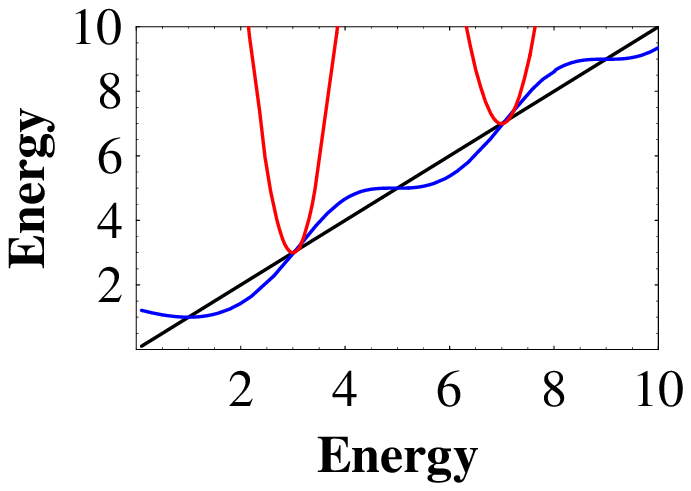}}}
\rotatebox{0.0}{\scalebox{1.}{\includegraphics{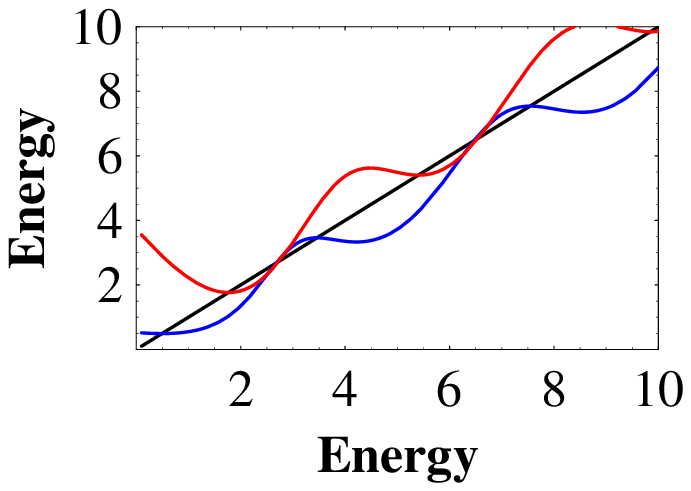}}}
\caption{\footnotesize{Left figure: harmonic oscillator. Right figure: quartic oscillator. The black line is the energy-line, the blue curve gives the energy-function of the even states, the red curve is the energy-function of the odd states. Crossed extrema define eigenvalues.}}
\label{1D_Harmonic_oscillator}\end{figure}

We study the Schr\"odinger-equation
\begin{equation}
-\,\frac{d^{2}}{dx^{2}}\,\psi(x)\,+\,a_{n}\,x^{n}\,\psi(x)\,=\,E\,\psi(x),\,n\,=\,(2,\,4),\, a_{2}\,=\,1,\,a_{4}\,=\,0.1\quad.
\end{equation}
The eigenvalues of the harmonic oscillator then are given by $E_{n} = 2 n + 1$. In Fig. (\ref{1D_Harmonic_oscillator}) we illustrate the energy-functions calculated by Eqs. (\ref{Energycondition1}, \ref{Energycondition2}). The blue curve is a result of the symmetric state $\phi_{\mathrm{s}}=1$, the red curve is a result of the anti-symmetric state $\phi_{\mathrm{odd}}=x$. The black line is the energy-line $E=E$. It can easily be deduced that the energy-line crosses the energy-functions, however, only the intersections that also fulfill the extremal-condition Eq. (\ref{Energycondition2}) are eigenvalues. This confirms that intersections define eigenvalues if, and only if also the extremal-condition is fulfilled in order to guarantee stability.
\begin{table}[t]
\begin{tabular}{c||c|c|c|c|c|c|}
\hline
 & $E_{0}$ & $E_{1}$ & $E_{2}$ & $E_{3}$ & $E_{4}$ & $E_{5}$ \\
\hline
even & 0.492174 & - & 3.46057 & - & 7.54779 & - \\
\hline
odd & - & 1.76363 & - & 5.40487 & - & 9.58751 \\
\hline
\end{tabular}
\caption{Eigenvalues of the quartic oscillator as illustrated in Fig. (\ref{1D_Harmonic_oscillator}).}
\label{Quartictable}\end{table}

The eigenvalues of the quartic oscillator are given by Tab. (\ref{Quartictable}). They are not equidistant and thus suggest a Maslov-index $\mu\,=\,\mu(E)$ as a function of the energy. We shall, however, not engage into a discussion about Maslov-indices here, since this topic is not the theme of this paper.

\subsection{Symmetric and and skew double-well potential}
\begin{figure}[t]\centering\vspace{-.05cm}
\rotatebox{0.0}{\scalebox{1.}{\includegraphics{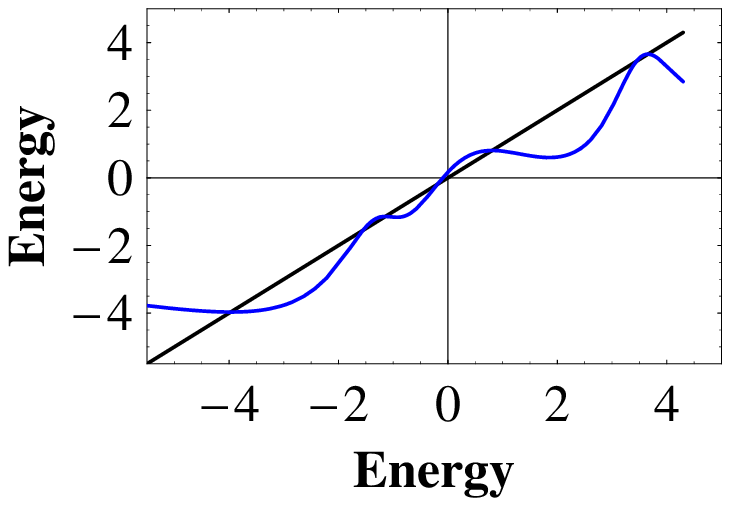}}}
\rotatebox{0.0}{\scalebox{1.}{\includegraphics{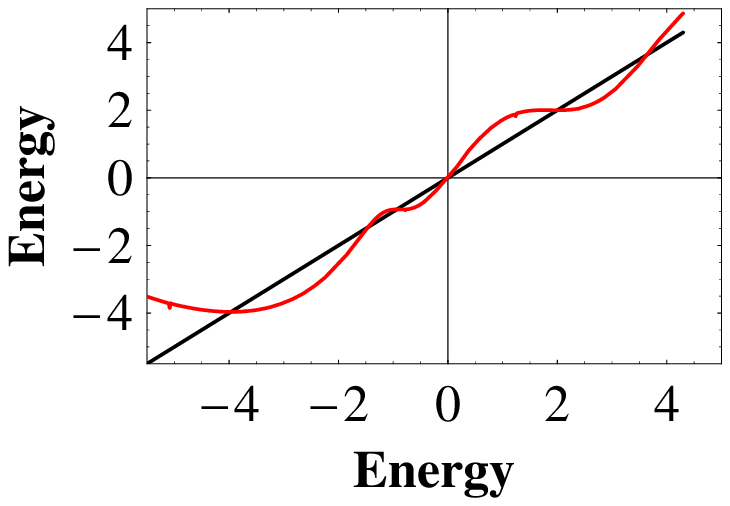}}}
\caption{\footnotesize{Symmetric double-well. Left figure: energy-line (black) and even energy-function (blue). Right figure: energy-line (black) and odd energy-function (red). Crossed extrema define eigenvalues.}}
\label{1D_Doublewell_even}\end{figure}

Interesting systems to study are multi-valley potentials, and potentials that are skew in some sense. In this subsection we will study a double-well potential that is given by the polynomial
\begin{equation}
U(x)\,=\,\frac{1}{10}\,x^{4}\,-\,a_{n}\,x^{3}\,-\,\frac{3}{2}\,x^{2},\,a_{n}\,=\,\left\{0,\,\frac{1}{10}\right\}\quad.
\label{doublewellpotential}\end{equation}
Depending on the value of $a_{n}$ the potential is a symmetrical double-well, or a skew double-well. First we turn to the symmetrical case. As it is illustrated in Fig. (\ref{1D_Doublewell_even}), we find that the ground-state and the first excited state are nearly degenerate. However, as the energy approaches threshold from below the levels spread, and above threshold the system behaves as one would usually expect it.

\begin{figure}[h]\centering\vspace{-.05cm}
\rotatebox{0.0}{\scalebox{1.}{\includegraphics{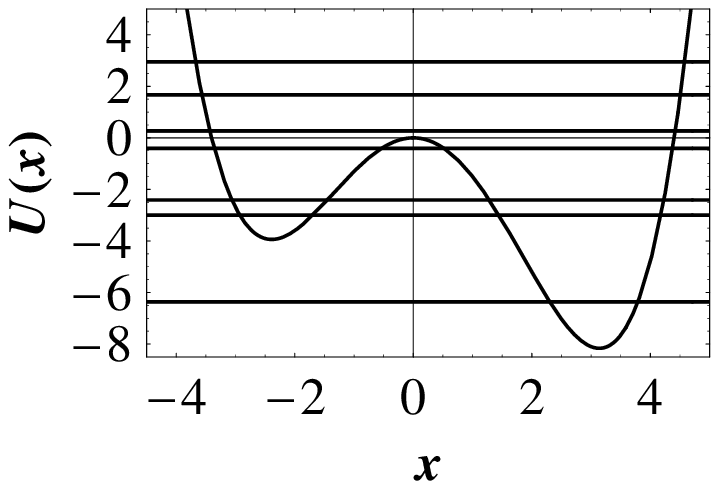}}}
\rotatebox{0.0}{\scalebox{1.}{\includegraphics{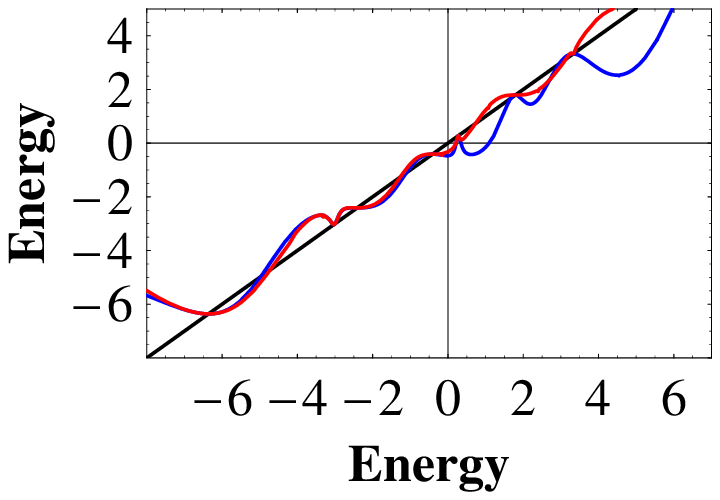}}}
\caption{\footnotesize{Left figure: Skew double-well potential. The horizontal lines denote eigenvalues. Right figure: Energy-curves of the skew double-well potential. The black line is the energy-line, the blue curve gives the energy-function of the even states, the red curve is the energy-function of the odd states. Crossed extrema define eigenvalues.}}
\label{1D_Doublewell_skew}\end{figure}

In the case of the skew double-well, see Fig. (\ref{1D_Doublewell_skew}) and Tab. (\ref{DoublewellSkew}), we find that the first four states $\{E_{0}, E_{1}, E_{2}, E_{3}\}$ are almost degenerate. However, the states $\{E_{4}, E_{5}, E_{6}, E_{7}\}$ are degenerate. The degeneracy can only be explained by the asymmetry of the potential and threshold effects. It seems to be difficult for the system to establish states of clearly defined symmetry as long as the skewness of the wells takes effect. Above threshold the skew influence declines, and we find states with clearly defined symmetry and thus non-degenerate eigenvalues.

It may, however, very well be that the apparent degeneracy is due to an insufficient numerical accuracy. This but seems not likely to us, since for different spacings of the lattice the results did not change.

\begin{table}[t]
\begin{tabular}{c||c|c|c|c|c|c|c|}
\hline
 & $E_{0}$ & $E_{1}$ & $E_{2}$ & $E_{3}$ & $E_{4}$ & $E_{5}$ & $E_{6}$ \\
\hline
even & $-3.96912$ & $-$ & $-1.14291$ & $-$ & 0.809548 & $-$ & 3.65765 \\
\hline
odd & $-$ & $-3.96484$ & $-$ & $-0.933845$ & $-$ & 1.99916 & $-$ \\
\hline
\end{tabular}
\caption{Eigenvalues of the symmetric double-well potential as illustrated in Fig. (\ref{1D_Doublewell_even}).}
\label{Doublewell}\end{table}
A note an a technical detail must be made. The potential Eq. (\ref{doublewellpotential}) has three extrema, and an odd state can be expected to have a node at each extremum. Thus, a natural ansatz for the odd state is given by $\phi_{\mathrm{odd}}(x)=(x-x_{1})(x-x_{2})(x-x_{3})$, where the $x_{i}$ denote the extrema. This ansatz works and yields the correct eigenvalues, but the resulting energy-curve shows large amplitudes and also tends to noisiness away from the eigenvalues, such that it is neither well-behaved, nor easy to work with. This is, of course, due to the polynomial structure of  $\phi_{\mathrm{odd}}(x)$. However, it is in fact enough to work with the local maximum, such that $\phi_{\mathrm{odd}}(x)=x$. We assume that this observation holds in general, since already one point of anti-symmetry is enough to define an anti-symmetric state.
\begin{table}[h!]
\begin{tabular}{c||c|c|c|c|c|c|c|c|c|c|c|c|c|}
\hline
 & $E_{0}$ & $E_{1}$ & $E_{2}$ & $E_{3}$ & $E_{4}$ & $E_{5}$ & $E_{6}$ & $ E_{7}$ & $E_{8}$ & $E_{9}$ & $E_{10}$ & $E_{11}$ & $E_{12}$ \\
\hline
even & $-6.35979$ & $-$ & $-2.99656$ & $-$ & $-2.41036$ & $-$ & $-0.4091$ & $-$ & 0.266 & $-$ & 1.6679 & $-$ & 2.9507 \\
\hline
odd & $-$ & $-6.35973$ & $-$ & $-2.99485$ & $-$ & $-2.41036$ & $-$ & $-0.4091$ & $-$ & 0.2763 & $-$ & 1.7931 & $-$  \\
\hline
\end{tabular}
\caption{Eigenvalues of the skew double-well potential as illustrated in Fig. (\ref{1D_Doublewell_skew}).}
\label{DoublewellSkew}\end{table}

\subsection{Potential with finite depth}
In systems with an attractive potential the \emph{incoming} state has to fulfill an additional condition, that is not related to the symmetry of space, but to a physical requirement. This is, that towards infinity the wave-function must decay like an exponential
\begin{equation}
\lim_{x\rightarrow\pm\infty}\psi(x)\,\sim\,\exp\left[-\,|x|\right]\quad.
\label{limitcondition}\end{equation}
As it turns out, it is indeed not necessary to make assumptions about the momentum of the decaying wave-function, because all information about the special properties of the decay is stored in the Lippmann-Schwinger equation. Thus, we understand that the \emph{incoming} state $\phi$ is just somewhat a skeleton, that describes the requirements of the symmetry and the physics in the most elementary way that is possible. Consequently, the \emph{incoming} states must read
\begin{equation}
\phi_{\mathrm{even}}(x)\,=\,\exp\left[-\,|x|\right],\,\phi_{\mathrm{odd}}(x)\,=\,x\,\exp\left[-\,|x|\right]\quad,
\label{initialcondition1}\end{equation}
such that the symmetry of space and the physical requirement of decay are combined.

We study the inverse, attractive cosh-potential
\begin{equation}
U(x)\,=\,-\frac{U_{0}}{\cosh[x]^{2}},\,U_{0}\,=\,25\quad.
\label{coshpotential}\end{equation}
The formula for the eigenvalues is known, see e.g. \cite{Landau},
\begin{equation}
E_{n}\,=\,-\,\frac{1}{4}\,\left(-\,(1\,+\,2\,n)\,+\,\sqrt{1\,+\,4\,U_{0}}\right)^{2}\quad,
\label{cosheigenvalues}\end{equation}
where we already have applied our present scaling.
\begin{figure}[t]\centering\vspace{-1.05cm}
\rotatebox{0.0}{\scalebox{1.0}{\includegraphics{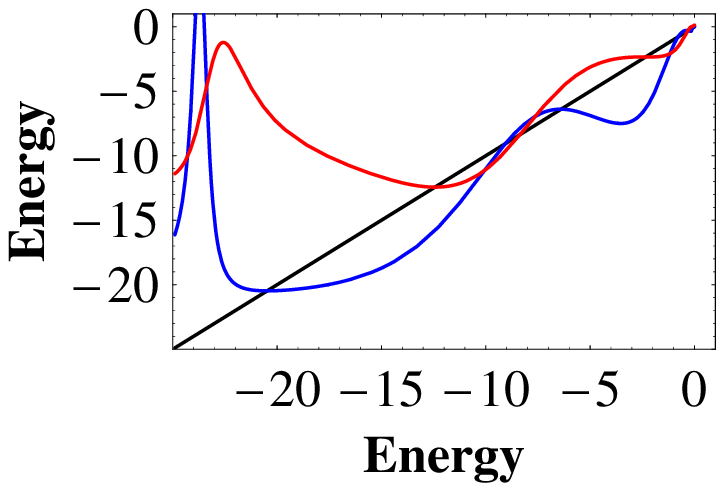}}}
\rotatebox{0.0}{\scalebox{1.}{\includegraphics{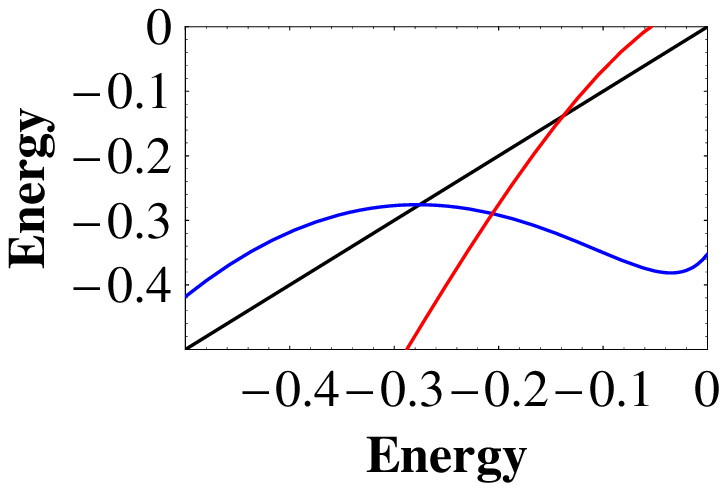}}}
\caption{\footnotesize{Left figure: Energy-curves of the inverse, attractive cosh-potential. The black line is the energy-line, the blue curve is the energy-function of the even states, the red curve is the energy-function for the odd states. Right figure: The same for the highest eigenstate just below threshold.}}
\label{1D_Cosh}\end{figure}
The energy-curves are given by Fig. (\ref{1D_Cosh}), and we see that the numerically calculated eigenvalues meet the values that are given by the formula Eq. (\ref{cosheigenvalues}). We encounter the well-known phenomenon that the distance of the eigenvalues declines towards threshold.  Fig. (\ref{1D_Cosh}) also shows that the highest eigenvalue just below threshold is given correctly.

\subsection{Singular potentials}
As an example for an attractive inverse power-law potential we study one-dimensional hydrogen, given by the Schr\"odinger-equation
\begin{equation}
-\,\frac{d^{2}}{dx^{2}}\,\psi(x)\,-\,\frac{1}{|x|}\,\psi(x)\,=\,E\,\psi(x)\quad.
\label{1Dhydrogen}\end{equation}
In our present scaling the eigenvalues are given by $E_{n}=-1/(4\,n^{2})$. All states are of odd symmetry, since it must hold that $\psi(x=0)=0$.
\begin{figure}[t]\centering\vspace{-1.05cm}
\rotatebox{0.0}{\scalebox{1.0}{\includegraphics{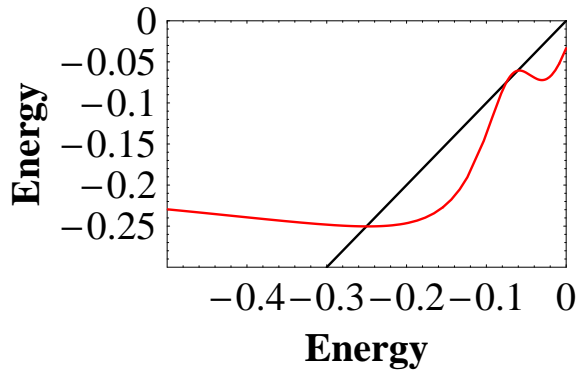}}}
\caption{\footnotesize{ Energy-curve of one-dimensional hydrogen. The black line is the energy-line, the red curve is the energy-function for the odd states. Shown are the ground-state and the first excited state.}}
\label{1D_hydrogen}\end{figure}
In Fig. (\ref{1D_hydrogen}) we illustrate the ground-state and the first excited state. In order to detect more states below threshold, it is of course clear that the numerical accuracy has to be increased from level to level in order to get reliable results. The numerical accuracy we have used to calculate the first two eigenvalues, as it can be seen from Fig. (\ref{1D_hydrogen}), is not high enough to achieve a more detailed resolution of the threshold. It seems that our method has reached it's limits here, since a higher resolution of the threshold would only be possible with an unreasonable numerical effort.

There is, however, the possibility to remove the singularity and thus to regularize the problem by the Duru-Kleinert transform, see e.g. \cite{Kleinert1,  Kleinert2}. We can cast the Schr\"odinger-equation Eq. (\ref{1Dhydrogen}) into
\begin{equation}
-\,\frac{d^{2}}{dy^{2}}\,\psi(y)\,+\,4\,\omega\,y^{2}\,\psi(y)\,=\,4\,\psi(y)\quad,
\label{DuruKleinert1}\end{equation}
where we have set $x=y^{2}$ and $\omega=-E$. The pseudo-energy $\epsilon$ is $\epsilon=4$, of course, such that the Greens-function must read $G(|x-x'|,\,\epsilon\,=\,4)$. From this follows, that the eigenvalues are determined by $\epsilon(\omega)=4$. The Coulomb-problem is thus mapped on a harmonic oscillator and is easy to treat.

We now turn to the problem of higher order singularities, given by the Schr\"odinger-equation
\begin{equation}
-\,\frac{d^{2}}{dx^{2}}\,\psi(x)\,-\,\frac{\beta_{\alpha}}{|x|^{\alpha}}\,\psi(x)\,=\,E\,\psi(x)\quad.
\label{higherorder1}\end{equation}
With the Duru-Kleinert regularization $|x|=\exp[|y|]$ we obtain
\begin{equation}
-\,\frac{d^{2}}{dy^{2}}\,\psi(y)\,-\,\beta_{\alpha}\,\exp[(2\,-\,\alpha)\,|y|]\,\psi(y)\,+\,\omega\,\exp[2\,|y|]\,\psi(y)\,=\,0\,,\quad\alpha\,\geq\,2\quad,
\label{higherorder1}\end{equation}
where we have set $\omega=-E$ as above. For the pseudo-energy $\epsilon(\omega)=0$ must hold. In the case of a van der Waals-potential tail we have $\alpha=3$ and chose $\beta_{3}=5$, from which we find a ground-state for $E_{0}=-\,0.0191353$\,. The procedure that allows us to calculate $E_{0}$ is explained in the next section.

\section{Eigenvalue-engineering}
In this section we demonstrate how our method can be used to prepare a state with a certain energy. We treat the case of a symmetric double-well potential
\begin{equation}
U(x)\,=\,\frac{1}{10}\,x^{4}\,-\,a\,|x|^{\gamma}\quad.
\label{engineering1}\end{equation}
As an example, we seek the numerical values of the parameters $\{a, \gamma\}$, such that the double-well potential has the even eigenvalue $E_{n}=-3$\,. Consequently, we must set $G(|x-x'|, E=-3)$ for the Greens-function. With the fixed Greens-function we calculate the energy as a parametric function
\begin{equation}
\epsilon(\beta)\,=\,\left<\psi(x,\,\beta)|\hat{H}(\beta)|\psi(x,\,\beta)\right>\quad.
\label{engineering2}\end{equation}
In Eq. (\ref{engineering2}) $\beta$ stands either for $a$ or for $\gamma$. A possible value for a parameter is given if $\epsilon(\beta)=-3$ holds. From above but we know that a crossing of the energy-line does not necessarily mean that an energy is an eigen-energy. Thus, we need an additional cross-check by the extremal-condition Eq. (\ref{Energycondition2}).
\begin{figure}[t!]\centering\vspace{-.05cm}
\rotatebox{0.0}{\scalebox{1.0}{\includegraphics{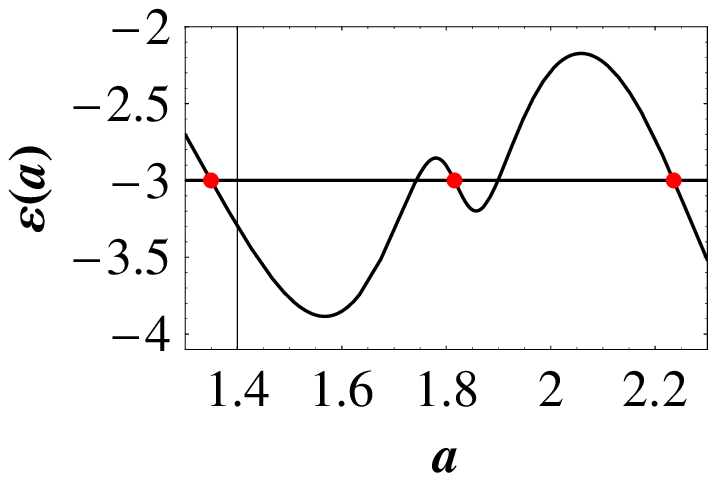}}}
\rotatebox{0.0}{\scalebox{1.}{\includegraphics{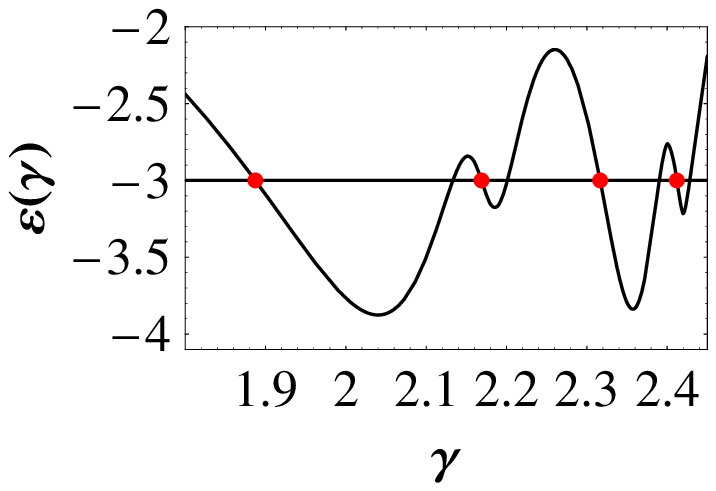}}}
\caption{\footnotesize{The red dots mark numerical values for even states. Left figure: Parametric energy-function $\epsilon(a)$ for the double-well potential Eq. (\ref{engineering1} with fixed $\gamma=2$\,. Right figure: The same for a fixed $a=3/2$\,.}}
\label{Engineering}\end{figure}

In Fig. (\ref{Engineering}) we illustrate the behaviour of the parametric energy-functions $\epsilon(a), \epsilon(\gamma)$ for even states.  Each crossing of the line $\epsilon=-3$ marks a possible eigenstate, but not every crossing is an eigenstate. Only the values that are marked with red dots also fulfill the extremal-condition given by Eq. (\ref{Energycondition2}). We observe, that an eigen-state is given if the crossing takes place from above. However, we are not able explain this behaviour in terms of a rule, since to us there is no reasonable argument for this behaviour.
The values for the parameters are given in Tab. (\ref{tabengineering}).
\begin{table}[h!]
\begin{tabular}{c||c|c|c|c|}
\hline
$a$ & 1.34938 &  1.81571 & 2.23559 & - \\
\hline
$\gamma$ & 1.8873 &  2.16885 & 2.31662 & 2.41205 \\
\hline
\end{tabular}
\caption{Numerical values for the parameters $\{a, \gamma\}$.}
\label{tabengineering}\end{table}
Note that the same crossing-behaviour can be observed in the search for eigenvalues of singular-potentials after the Duru-Kleinert transform has been carried out.

\section{Conclusion}
We have reported a method to calculate eigenvalues of a quantum system that, as far as we know, is different from other approaches in this field. Our method makes use of the basic properties of the Schr\"odinger-equation in terms of it's Greens-function and it's integral-representation, which is the Lippmann-Schwinger equation. In particular, we exploit the fact that the Lippmann-Schwinger equation can be solved exactly by matrix-inversion. As we have illustrated in the figures, our approach is similar to the shooting-method in the sense that the eigenvalues are determined by a systematic search-algorithm. Once programmed, our approach can be applied in general in terms of plug and play.

Furthermore, we use the symmetry of space, described by the solutions of the Laplace-equation, to define \emph{incoming} states that serve as basic input for the Lippmann-Schwinger equation. The \emph{incoming} states act as projectors for even and odd wave-functions, such that the eigenvalue-problem can be treated analogously to the scattering-problem.

We have demonstrated that our method can successfully be applied on a variety of potentials: polynomial potentials like the harmonic and the quartic oscillator, and double-well potentials. Furthermore, we have treated attractive potentials with finite and infinite depth. Finally, we have illustrated how our method can be used for eigenvalue-engineering.

The use of the Greens-function also makes it possible to extend our method to higher dimensions. Especially, the use of the Greens-function will allow to treat non-separable problems.

All our calculations have been carried out by ordinary Riemann-sums with equidistant intervals. Higher numerical accuracy may still be achieved when more sophisticated methods of integration are applied. This, however, together with the treatment of higher dimensions, shall be left for further work. The focus of our present paper is set on the principal demonstration of our method.

\end{document}